\documentclass[prd,twocolumn,groupedaddress,showpacs,nofootinbib]{revtex4}
\usepackage{graphicx}
\usepackage{dcolumn}
\usepackage{amssymb}
\usepackage{mathrsfs}
\usepackage{amsmath}
\usepackage{epsfig}
\usepackage[dvips]{color}
\usepackage{hhline}

\begin{document}

\title{Inflationary baryogenesis with low reheating temperature and testable neutron-antineutron oscillation }

\author{Pei-Hong Gu$^{1,2}_{}$}
\email{peihong.gu@mpi-hd.mpg.de}

\author{Utpal Sarkar${}^{3,4}$}
\email{utpal@prl.res.in}

\affiliation{ ${}^1$Max-Planck-Institut f\"{u}r Kernphysik,
Saupfercheckweg 1,
69117 Heidelberg, Germany\\
${}^2$Institute of Advanced Studies, Nanyang Technological
University,
Singapore 639673\\
${}^3$Physical Research Laboratory, Ahmedabad 380009, India\\
${}^4$McDonnell Center for the Space Sciences, Washington
University, St. Louis, MO 63130, USA}

\begin{abstract}

Recently we extended the standard model by four TeV-scale fields
including a singlet fermion, an isotriplet and two isosinglet
diquark scalars to generate the cosmological baryon asymmetry with
an observable neutron-antineutron oscillation. We now
supersymmetrize our model but do not constrain it at the TeV scale.
The superpartner of the singlet fermion can serve as an inflaton
field. Its three-body decays, mediated by the isosinglet diquarks
and their superpartners, can simultaneously provide a low reheating
temperature and a sizable CP asymmetry. We thus can realize a
nonthermal baryogenesis without the gravitino problem. Meanwhile, we
can have a testable neutron-antineutron oscillation induced by the
exchange of one isosinglet and two isotriplet diquarks if the
isotriplet diquark is at the TeV scale.

\end{abstract}


\maketitle

Many baryogenesis scenarios
\cite{yoshimura1978,ad1985,fy1986,ck1987,cr1991,mz1992,fy2002} have
been studied to explain the observed baryon asymmetry in the present
universe. A successful baryogenesis model should respect the
Sakharov conditions \cite{sakharov1967} (baryon number
nonconservation, C and CP violation as well as departure from
equilibrium) unless CPT is not invariant \cite{ck1987}. If the
baryogenesis works above the weak scale, the baryon number violation
should be modified to be a baryon-minus-lepton $(B-L)$ number
violation in the presence of the sphalerons \cite{krs1985}, which
violate the baryon and lepton numbers by an equal amount. The
sphaleron processes will not affect any primordial $B-L$ asymmetry
and will convert the $B-L$ asymmetry to a baryon asymmetry and a
lepton asymmetry. The $B-L$ asymmetry can be composed of a pure
baryon asymmetry or a pure lepton asymmetry or any unequal baryon
and lepton asymmetries. The baryon and/or lepton number violation
can lead to other interesting phenomenologies. For example, the
baryon number violation of two units could result in a
neutron-antineutron oscillation \cite{mm1980}, the lepton number
violation of two units could result in a neutrinoless double beta
decay \cite{furry1939}, whereas the combined baryon and lepton
number violation could result in a proton decay \cite{ps1973}.

In a recent work \cite{gs2011}, we extended the $SU(3)_c^{}\times
SU(2)_L^{}\times U(1)_Y^{}$ standard model (SM) by four TeV-scale
fields (a singlet fermion, an isotriplet and two isosinglet diquark
scalars) to generate the cosmological baryon asymmetry with an
observable neutron-antineutron oscillation. In this paper we will
study the supersymmetric version of our model but will not constrain
it at the TeV scale. Like the sneutrinos (the superpartner of the
right-handed neutrinos in the supersymmetric seesaw
\cite{minkowski1977} model), the superpartner of the singlet fermion
can drive a chaotic inflation \cite{msyy1993}. Through the
three-body decays \cite{hambye2001} of the inflaton, the universe
can be reheated at a low temperature to avoid the gravitino problem
\cite{kl1983}. At the same time, we can obtain a sizable CP
asymmetry in the inflaton decays to realize a nonthermal
baryogenesis. Furthermore, we can have a neutron-antineutron
oscillation by the exchange of one isosinglet diquark and two
isotriplet diquarks. This neutron-antineutron oscillation can be
sensitive to the future experiments if the isotriplet diquark is at
the TeV scale.

For simplicity, we only write down the superpotential relevant to
our demonstration,
\begin{eqnarray}
\label{super} \mathcal{W}&\supset& y_{ai}^{} \hat{\Delta}_a^{}
\hat{U}^c_i\hat{X}^c_{} + \bar{f}_{aij}^{}
\hat{\bar{\Delta}}_a^{}\hat{D}_i^c \hat{D}_j^c + h_{ij}^{}
\hat{Q}_i^{} i
\tau_2^{} \hat{\Omega} \hat{Q}_j^{}  \nonumber\\
&& + \kappa_a^{} \hat{\Delta}_a^{}
\textrm{Tr}(\hat{\Omega}\hat{\Omega})+ \bar{\kappa}_a^{}
\hat{\bar{\Delta}}_a^{}
\textrm{Tr}(\hat{\bar{\Omega}}\hat{\bar{\Omega}})+\frac{1}{2}M_X^{}\hat{X}^c_{}
\hat{X}^c_{}\nonumber\\
&&+M_a^{} \hat{\bar{\Delta}}_a^{}\hat{\Delta}_a^{}+M_\Omega^{}
\textrm{Tr}(\hat{\bar{\Omega}}\hat{\Omega})\,.
\end{eqnarray}
Here
\begin{eqnarray}
\hat{X}(\textbf{1},\textbf{1},0)
\end{eqnarray}
is a singlet superfield,
\begin{eqnarray}
\begin{array}{ll}
\hat{\Delta}(\textbf{3},\textbf{1},+\frac{2}{3})\,,&
\hat{\Omega}(\textbf{3},\textbf{3},-\frac{1}{3})=\left[
\begin{array}{rr}
\frac{1}{\sqrt{2}}\hat{\omega}_{\frac{1}{3}}^{}\quad\quad&\hat{\omega}_{\frac{2}{3}}^{}\\
[3mm]
\hat{\omega}_{\frac{4}{3}}^{}\quad\quad&-\frac{1}{\sqrt{2}}\hat{\omega}_{\frac{1}{3}}^{}
\end{array}\right]\,,\\
[7mm] \hat{\bar{\Delta}}(\textbf{3},\textbf{1},-\frac{2}{3})\,, &
\hat{\bar{\Omega}}(\textbf{3},\textbf{3},+\frac{1}{3})=\left[
\begin{array}{rr}
\frac{1}{\sqrt{2}}\hat{\bar{\omega}}_{\frac{1}{3}}^{}\quad\quad&\hat{\bar{\omega}}_{\frac{4}{3}}^{}\\
[3mm]
\hat{\bar{\omega}}_{\frac{2}{3}}^{}\quad\quad&-\frac{1}{\sqrt{2}}\hat{\bar{\omega}}_{\frac{1}{3}}^{}
\end{array}\right]
\end{array}
\end{eqnarray}
are the diquark superfields, while
\begin{eqnarray}
\hat{Q}(\textbf{3},\textbf{2},\frac{1}{6})\,,\quad
\hat{U}(\textbf{3},\textbf{1},\frac{2}{3})\,,\quad
\hat{D}(\textbf{3},\textbf{1},-\frac{1}{3})
\end{eqnarray}
denote the usual quark superfields. Corresponding to the baryon
number $B=\frac{1}{3}$ of the quark superfields, we assign the
baryon number $B=-1$ for the singlet superfield while the baryon
number $B=-\frac{2}{3}$ for the diquark superfields
$(\hat{\Delta},\hat{\bar{\Delta}}^c_{})$ and $(\hat{\Omega},
\hat{\bar{\Omega}}^c_{})$. The singlet superfield is forbidden to
have the couplings with the lepton and Higgs superfields so that we
can avoid the dangerous proton decay. For this purpose, we can
introduce certain discrete symmetries, such as a $Z_2^{}$ symmetry
under which the lepton superfields (including the right-handed
neutrino superfields for the seesaw mechanism) are odd while the
others are even. The superpotential will yield a lagrangian as
below,
\begin{eqnarray}
\label{lagrangian} \mathcal{L}&\supset&
-y_{ai}^{}(\delta_a^{}\bar{u}_{Ri}^{}X_R^c+\tilde{X}^\ast_{}\bar{u}_{Ri}^{}\tilde{\delta}_{La}^{}
+\tilde{u}_{Ri}^\ast\bar{X}_R^{}\tilde{\delta}_{La}^{}\nonumber\\
&&+M_a^{}\bar{\delta}_a^\ast \tilde{u}_{Ri}^\ast \tilde{X}^\ast_{})-
\bar{f}_{aij}^{}(\bar{\delta}_a^{}\bar{d}_{Ri}^{}d_{Rj}^c
+2\bar{d}_{Ri}\tilde{\bar{\delta}}_{La}^{}\tilde{d}_{Rj}^\ast\nonumber\\
&&+M_a^{}\delta_a^\ast \tilde{d}_{Ri}^\ast
\tilde{d}_{Rj}^\ast)-h_{ij}^{}(\bar{q}_{Li}^c i \tau_2^{} \omega
q_{Lj}^{}+2\bar{q}_{Li}^c i \tau_2^{} \tilde{\omega}_L^{}
\tilde{q}_{Lj}^{}\nonumber\\
&&+M_\Omega^{}\tilde{q}_{Li}^T i \tau_2^{} \bar{\omega}^\dagger_{}
\tilde{q}_{Lj}^{})-\kappa_a^{}[\delta_a^{}\textrm{Tr}(\bar{\tilde{\omega}}_L^c\tilde{\omega}_L^{})\nonumber\\
&& +2\textrm{Tr}(\bar{\tilde{\omega}}_L^c
\omega)\tilde{\delta}_{La}^{}+\bar{\delta}_a^\ast\textrm{Tr}(\omega^2_{})]
-\bar{\kappa}_a^{}[\bar{\delta}_a^{}\textrm{Tr}(\bar{\tilde{\bar{\omega}}}_L^c\tilde{\bar{\omega}}_L^{})\nonumber\\
&& +2\textrm{Tr}(\bar{\tilde{\bar{\omega}}}_L^c
\bar{\omega})\tilde{\bar{\delta}}_{La}^{}+\delta_a^\ast\textrm{Tr}(\bar{\omega}^2_{})]
-\frac{1}{2}M_X^{}\bar{X}_R^{} X_R^c \nonumber\\
&&-M_a^{}
\bar{\tilde{\delta}}_{La}^c\tilde{\bar{\delta}}_{La}^{}-M_\Omega^{}
\textrm{Tr}(\bar{\tilde{\omega}}_{L}^c\tilde{\bar{\omega}}_L^{})+\textrm{H.c.}-M_X^2\tilde{X}^\ast_{}
\tilde{X}\nonumber\\
&&-M_a^2(\delta_a^\ast\delta_a^{}+\bar{\delta}_a^\ast\bar{\delta}_a^{})
-M_\Omega^2[\textrm{Tr}(\omega_{}^\dagger\omega)\nonumber\\
&&+\textrm{Tr}(\bar{\omega}_{}^\dagger\bar{\omega})]\,.
\end{eqnarray}

The singlet scalar $\tilde{X}$ can drive a chaotic inflation, like
the sneutrino in the supersymmetric seesaw model \cite{msyy1993}.
Specifically, the inflaton $\tilde{X}$ will begin to oscillate at
the time $t\sim 1/H(T)\sim 1/M_X^{}$. Here the Hubble constant $H$
is given by
\begin{eqnarray}
H(T)=\left(\frac{8\pi^{3}_{}g_{\ast}^{}}{90}\right)^{\frac{1}{2}}_{}
\frac{T^{2}_{}}{M_{\textrm{Pl}}^{}}
\end{eqnarray}
with $M_{\textrm{Pl}}^{}=\mathcal{O} (10^{19}_{}\,\textrm{GeV})$
being the Planck mass and $g_\ast^{}=\mathcal{O}(200)$ being the
relativistic degrees of freedom. Subsequently, the inflaton will
start to decay at the time $t\sim 1/H(T)\sim 1/\tilde{\Gamma}$ with
$\tilde{\Gamma}$ being the decay width. The universe then can be
reheated by the relativistic decay products. The reheating
temperature is determined by
\begin{eqnarray}
\tilde{\Gamma}=H(T)\Rightarrow
T_R^{}=\left(\frac{90}{8\pi^{3}_{}g_{\ast}^{}}\right)^{\frac{1}{4}}_{}
\sqrt{M_{\textrm{Pl}}^{}\tilde{\Gamma}}\,.
\end{eqnarray}

\begin{figure*}
\vspace{9.8cm} \epsfig{file=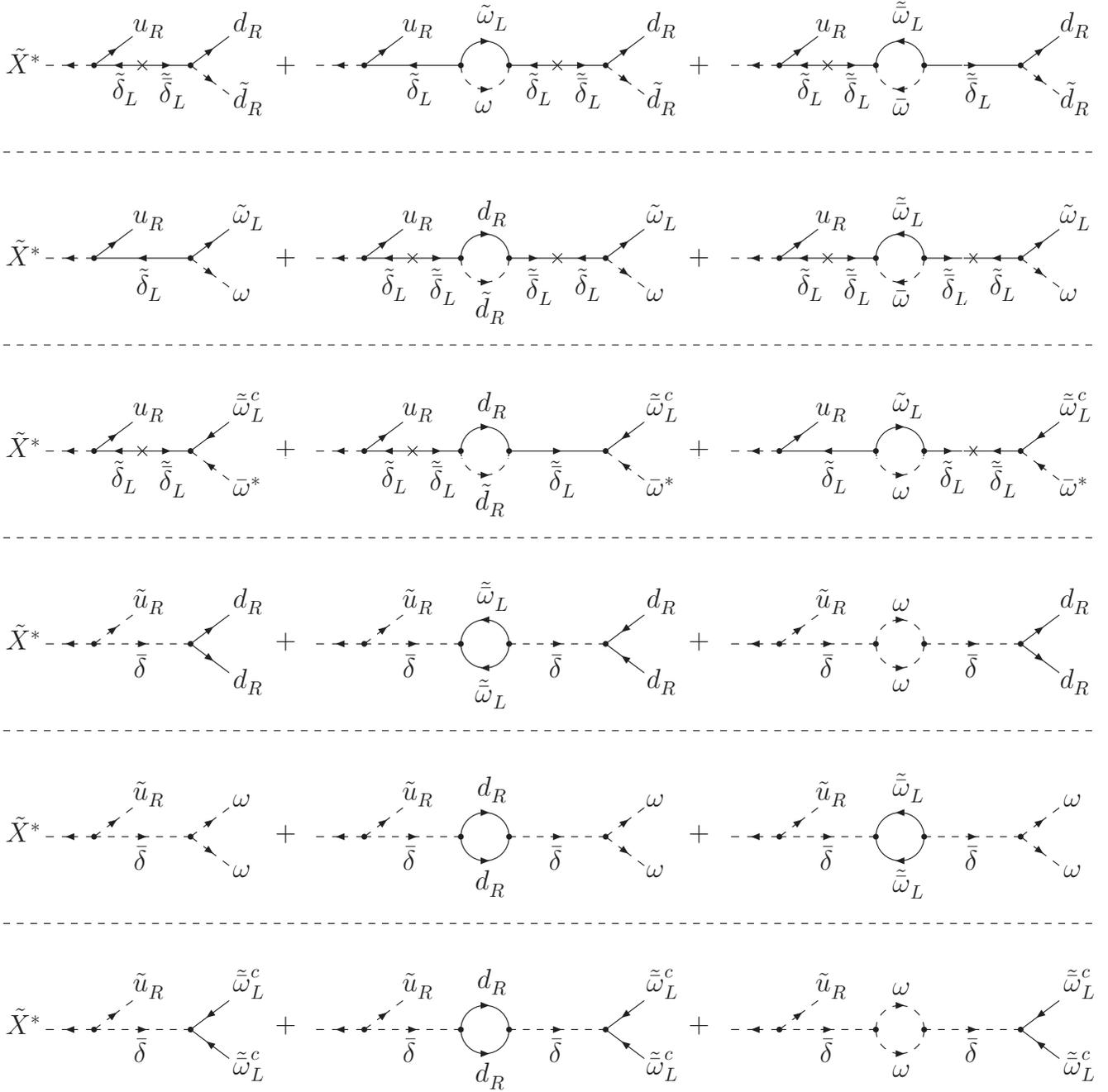, bbllx=5.5cm, bblly=6.0cm,
bburx=15.5cm, bbury=16cm, width=8.8cm, height=8.8cm, angle=0,
clip=0} \vspace{-0.7cm} \caption{\label{decay} The three-body decays
of the inflaton at tree level and one-loop order. The CP conjugation
is not shown for simplicity.}
\end{figure*}

The masses in the superpotential (\ref{super}) and the lagrangian
(\ref{lagrangian}) are assumed to hold the following hierarchy,
\begin{eqnarray}
2 M_\Omega^{} < M_X^{} < M_a^{}\,.
\end{eqnarray}
This means that the inflaton can only have the three-body decays
mediated by the isosinglet diquark scalars and their superpartners.
We show the inflaton decays at tree level and one-loop order in Fig.
\ref{decay}. The final states of the inflaton decays carry a baryon
number $B=+1$ or $ B=-1$, i.e.
\begin{eqnarray}
\begin{array}{ll}
~~~B =+1 & \quad\quad ~~~ B =-1 \\
[5mm] \tilde{X}^\ast_{}\rightarrow
u_R^{}d_R^{}\tilde{d}_R^{}\,,&\quad\quad \tilde{X}~\,\rightarrow
u_R^c d_R^c \tilde{d}_R^\ast\,,
\\
[3mm] \tilde{X}~\,\rightarrow u_R^c\tilde{\omega}_L^c
\omega^\ast_{}\,,&\quad\quad
 \tilde{X}^\ast_{}\rightarrow
u_R^{} \tilde{\omega}_L^{}\omega\,,\\
[3mm] \tilde{X}~\,\rightarrow
u_R^c\tilde{\bar{\omega}}_L^{}\bar{\omega}\,,
&\quad\quad\tilde{X}^\ast_{}\rightarrow u_R^{}\tilde{\bar{\omega}}_L^c\omega^\ast_{}\,,\\
[3mm] \tilde{X}^\ast_{}\rightarrow
\tilde{u}_R^{}d_R^{}d_R^{}\,,&\quad\quad \tilde{X}~\,\rightarrow
\tilde{u}_R^\ast d_R^c d_R^c\,,\\
[3mm] \tilde{X}~\,\rightarrow
\tilde{u}_R^\ast\omega^\ast_{}\omega^\ast_{}\,,
&\quad\quad\tilde{X}^\ast_{}\rightarrow \tilde{u}_R^{}\omega\omega\,,\\
[3mm] \tilde{X}~\,\rightarrow \tilde{u}_R^\ast
\tilde{\bar{\omega}}_L^{}\tilde{\bar{\omega}}_L^{}\,,
&\quad\quad\tilde{X}^\ast_{}\rightarrow
\tilde{u}_R^{}\tilde{\bar{\omega}}_L^c\tilde{\bar{\omega}}_L^c\,.
\end{array}
\end{eqnarray}
The $B=+1$ processes will generate a positive baryon number while
the $B=-1$ processes will generate a negative baryon number.
Therefore, the $B=+1$ processes and the $B=-1$ processes can have
different decay widths to induce a net baryon number if CP is not
conserved. In addition to the three-body decays, there are
$3\leftrightarrow3$ and $2\leftrightarrow 2$ scattering processes
violating the baryon number. As long as the reheating temperature is
far below the inflaton mass, the scattering processes will not wash
out the induced baryon asymmetry since they have completely
decoupled before the inflaton decays. We further assume that other
baryon or lepton number violating interactions does not exist or
have already decoupled. The baryon asymmetry from the inflaton
decays then should be \cite{msyy1993}
\begin{eqnarray}
\label{basymmetry} \frac{n_B^{}}{s}=\tilde{\varepsilon}
\frac{n_{\tilde{X}}^{}}{s}\left|_{T=T_R^{}}^{}\right.\simeq
\frac{3}{4} \tilde{\varepsilon}\frac{T_{R}^{}}{M_X^{}}\,.
\end{eqnarray}
Through the sphalreons, we can obtain a final baryon asymmetry
\cite{ks1988},
\begin{eqnarray}
\label{shpaleron} \eta_B^{}=\frac{8}{23}\frac{n_B^{}}{s}\,.
\end{eqnarray}

\begin{figure*}
\vspace{7.5cm} \epsfig{file=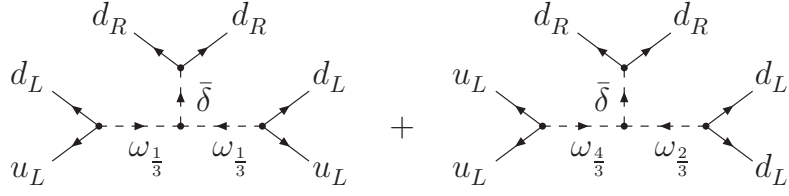, bbllx=4.5cm, bblly=6.0cm,
bburx=14.5cm, bbury=16cm, width=8.8cm, height=8.8cm, angle=0,
clip=0} \vspace{-12.5cm} \caption{\label{bviolation} The six-quark
interactions violating the baryon number of two units. The CP
conjugation is not shown for simplicity.}
\end{figure*}

The final baryon asymmetry is determined by the reheating
temperature and the CP asymmetry for a given inflaton mass. We hence
calculate the decay width at tree level and the CP asymmetry at
one-loop order,
\begin{widetext}
\begin{eqnarray}
\tilde{\Gamma}&=&\Gamma_{\tilde{X}}^{}=\Gamma_{\tilde{X}\rightarrow
u_R^c d_R^c \tilde{d}_R^\ast}^{}+\Gamma_{\tilde{X}\rightarrow u_R^c
\tilde{\omega}_L^c\omega^\ast_{}}^{} +\Gamma_{\tilde{X}\rightarrow
u_R^c\tilde{\bar{\omega}}_L^{}\omega}^{}+
\Gamma_{\tilde{X}\rightarrow \tilde{u}_R^\ast d_R^c d_R^c}^{}
+\Gamma_{\tilde{X}\rightarrow
\tilde{u}_R^\ast\omega^\ast_{}\omega^\ast_{}}^{}
+\Gamma_{\tilde{X}\rightarrow \tilde{u}_R^\ast
\tilde{\bar{\omega}}_L^{}\tilde{\bar{\omega}}_L^{}}^{}\nonumber\\
[2mm]
&=&\Gamma_{\tilde{X}^\ast_{}}^{}=\Gamma_{\tilde{X}^\ast_{}\rightarrow
u_R^{} d_R^{}
\tilde{d}_R^{}}^{}+\Gamma_{\tilde{X}^\ast_{}\rightarrow u_R^{}
\tilde{\omega}_L^{}\omega}^{} +\Gamma_{\tilde{X}^\ast_{}\rightarrow
u_R^{}\tilde{\bar{\omega}}_L^c\omega^\ast_{}}^{}+
\Gamma_{\tilde{X}^\ast_{}\rightarrow \tilde{u}_R^{} d_R^{}
d_R^{}}^{} +\Gamma_{\tilde{X}^\ast_{}\rightarrow
\tilde{u}_R^{}\omega\omega}^{} +\Gamma_{\tilde{X}^\ast_{}\rightarrow
\tilde{u}_R^{}
\tilde{\bar{\omega}}_L^c\tilde{\bar{\omega}}_L^c}^{}\nonumber\\
[2mm] &=&\frac{3}{2^9_{}\pi^3_{}}
\sum_{ab}^{}\left(\sum_{k}^{}y_{ak}^{}y_{bk}^\ast\right)
\left[2\left(\sum_{ij}^{}\bar{f}_{aij}^{}\bar{f}_{bij}^\ast+3\bar{\kappa}_a^{}\bar{\kappa}_b^\ast\right)r_a^{}r_b^{}
+\kappa_a^\ast\kappa_b^{}\left(6+r_a^2 r_b^2 \right)\right]\,,\\
[3mm]
\tilde{\varepsilon}&=&\frac{\Gamma_{\tilde{X}^\ast_{}\rightarrow
u_R^{}d_R^{}\tilde{d}_R^{}}^{}+\Gamma_{\tilde{X}\rightarrow u_R^c
\tilde{\omega}_L^c\omega^\ast_{}}^{} +\Gamma_{\tilde{X}\rightarrow
u_R^c\tilde{\bar{\omega}}_L^{}\omega}^{}
+\Gamma_{\tilde{X}^\ast_{}\rightarrow \tilde{u}_R^{}d_R^{}d_R^{}}^{}
+\Gamma_{\tilde{X}\rightarrow
\tilde{u}_R^\ast\omega^\ast_{}\omega^\ast_{}}^{}
+\Gamma_{\tilde{X}\rightarrow \tilde{u}_R^\ast
\tilde{\bar{\omega}}_L^{}\tilde{\bar{\omega}}_L^{}}^{}-\textrm{CP~conjugation}}
{\Gamma_{\tilde{X}^\ast_{}\rightarrow
u_R^{}d_R^{}\tilde{d}_R^{}}^{}+\Gamma_{\tilde{X}\rightarrow u_R^c
\tilde{\omega}_L^c\omega^\ast_{}}^{} +\Gamma_{\tilde{X}\rightarrow
u_R^c\tilde{\bar{\omega}}_L^{}\omega}^{}
+\Gamma_{\tilde{X}^\ast_{}\rightarrow \tilde{u}_R^{}d_R^{}d_R^{}}^{}
+\Gamma_{\tilde{X}\rightarrow
\tilde{u}_R^\ast\omega^\ast_{}\omega^\ast_{}}^{}
+\Gamma_{\tilde{X}\rightarrow \tilde{u}_R^\ast
\tilde{\bar{\omega}}_L^{}\tilde{\bar{\omega}}_L^{}}^{}+\textrm{CP~conjugation}}\nonumber\\
[2mm]
&=&\frac{1}{2\pi}\frac{\textrm{Im}\left[\sum_{abc}^{}\left(\sum_{k}^{}y_{bk}^\ast
y_{ck}^{}\right) \kappa_c^\ast \kappa_a^{}
\left(\sum_{ij}^{}\bar{f}_{aij}^{}\bar{f}_{bij}^\ast+3\bar{\kappa}_a^{}\bar{\kappa}_b^\ast\right)
\left(2-r_c^2\right)r_a^{}r_b^{}\right]}
{\sum_{ab}^{}\left(\sum_k^{}y_{ak}^{}y_{bk}^\ast\right)
\left[2\left(\sum_{ij}^{}\bar{f}_{aij}^{}\bar{f}_{bij}^\ast+3\bar{\kappa}_a^{}\bar{\kappa}_b^\ast\right)r_a^{}r_b^{}
+\kappa_a^\ast\kappa_b^{}\left(6+r_a^2 r_b^2 \right)\right]}\,,
\end{eqnarray}
\end{widetext}
where the parameter
\begin{eqnarray}
r_a^{}=\frac{M_X^{}}{M_a^{}}
\end{eqnarray}
has been defined. We further specify the decay width and the CP
asymmetry by
\begin{eqnarray}
\tilde{\Gamma}=\frac{3}{2^9_{}\pi^3_{}}A M_X^{}\,,\quad
\tilde{\varepsilon}=\frac{1}{2\pi}\frac{B}{A}\,,
\end{eqnarray}
where the quantities
\begin{widetext}
\begin{eqnarray}
A&=&\sum_k^{}y'^{2}_{1k}\left[2\left(\sum_{ij}^{}\bar{f}'^2_{1ij}+3\bar{\kappa}'^2_1\right)r_1^2
+\kappa'^2_1\left(6+r_1^4\right)\right]+
\sum_k^{}y'^{2}_{2k}\left[2\left(\sum_{ij}^{}\bar{f}'^2_{2ij}+3\bar{\kappa}'^2_2\right)r_2^2
+\kappa'^2_2\left(6+r_2^4\right)\right]\nonumber\\
&&+2\sum_{k}y'^{}_{1k}y'^{}_{2k}\left\{2\left[\sum_{ij}^{}\bar{f}'^{}_{1ij}\bar{f}'^{}_{2ij}
\cos\left(\alpha_k^{}+\bar{\beta}_{ij}^{}\right)
+3\bar{\kappa}'^{}_1\bar{\kappa}'^{}_2\cos\left(\alpha_k^{}+\bar{\gamma}\right)\right]r_1^{}r_2^{}
+\kappa'^{}_1\kappa'^{}_2\cos\left(\alpha_k^{}-\gamma\right)\left(6+r_1^2
r_2^2\right)\right\}\,,\\
[3mm] B&=&\sum_k^{}y'^{}_{1k}y'^{}_{2k}\kappa'^{}_1\kappa'^{}_2
\left[\left(\sum_{ij}^{}\bar{f}'^2_{1ij}+3\bar{\kappa}'^2_1\right)\left(2-r_2^2\right)r_1^2
-\left(\bar{f}'^2_{2ij}+3\bar{\kappa}'^2_2\right)\left(2-r_1^2\right)r_2^2\right]
\sin\left(\alpha_k^{}-\gamma\right)\nonumber\\
&&+\sum_k^{}y'^{}_{1k}y'^{}_{2k}\left[\sum_{ij}^{}\bar{f}'^{}_{1ij}\bar{f}'^{}_{2ij}
\sin\left(\alpha_k^{}+\bar{\beta}_{ij}^{}\right)
+3\bar{\kappa}'^{}_1\bar{\kappa}'^{}_2\sin\left(\alpha_k^{}+\bar{\gamma}\right)\right]\left[\kappa'^2_2\left(2-r_2^2\right)
-\kappa'^2_1\left(2-r_1^2\right)\right]r_1^{}r_2^{}\nonumber\\
&&+\kappa'^{}_{1}\kappa'^{}_{2}\left[\sum_{ij}^{}\bar{f}'^{}_{1ij}\bar{f}'^{}_{2ij}
\sin\left(\bar{\beta}_{ij}^{}+\gamma\right)+
3\bar{\kappa}'^{}_1\bar{\kappa}'^{}_2\sin\left(\gamma+\bar{\gamma}\right)\right]\left\{\sum_k^{}\left[y'^2_{1k}\left(2-r_1^2\right)
-y'^2_{2k}\left(2-r_2^2\right)\right]\right\}r_1^{}r_2^{}\,,
\end{eqnarray}
\end{widetext}
are determined by the parameters,
\begin{eqnarray}
\begin{array}{ll}
~y_{ak}^{}=y'^{}_{ak}e^{i\alpha_{ak}^{}}_{}\,,&
~\,\alpha_{k}^{}=\alpha_{1k}^{}-\alpha_{2k}^{}\,,\\
[2mm]
\,\bar{f}_{aij}^{}=\bar{f}'^{}_{aij}e^{i\bar{\beta}_{aij}^{}}_{}\,,&
~\bar{\beta}_{ij}^{}=\bar{\beta}_{1ij}^{}-\bar{\beta}_{2ij}^{}\,,\\
[2mm] ~~\,\kappa_a^{}=\kappa'^{}_a e^{i\gamma_a^{}}_{}\,,&
~~~\,\gamma=\gamma_{1}^{}-\gamma_{2}^{}\,,\\
[2mm] ~~\,\bar{\kappa}_a^{}=\bar{\kappa}'^{}_a
e^{i\bar{\gamma}_a^{}}_{}\,,&
~~~\,\bar{\gamma}=\bar{\gamma}_{1}^{}-\bar{\gamma}_{2}^{}\,.
\end{array}
\end{eqnarray}

We now indicate that our model can simultaneously provide a desired
baryon asymmetry and a low reheating temperature. For example, we
take
\begin{eqnarray}
&&y'^{}_{1k}=y'^{}_{2k}=y'\,,~\bar{f}'^{}_{1ij}=\bar{f}'^{}_{2ij}=\bar{f}'\,,
\nonumber\\
&&
\kappa'^{}_1=\frac{1}{2}\kappa'^{}_2=\kappa'\,,~\bar{\kappa}'^{}_1=\bar{\kappa}'^{}_2=\bar{\kappa}'\,,\nonumber\\
&&\alpha_{k}^{}=\bar{\beta}_{ij}^{}=\gamma=\bar{\gamma}=\frac{1}{2}\delta\,,~r_1^{}=r_2^{}=r\,,
\end{eqnarray}
to derive
\begin{eqnarray}
T_R^{}&=&\frac{9}{32\pi^2_{}}\left(\frac{5}{\pi
g_\ast^{}}\right)^{\frac{1}{4}}_{}
y'\left\{\left[4\left(3\bar{f}'^2_{}+\bar{\kappa}'^2_{}\right)r^2_{}\left(1+\cos\delta\right)\right.\right.\nonumber\\
&&\left.\left.+ 3\kappa'^2_{}\left(6+r^4_{}\right)
\right]M_X^{}M_{\textrm{Pl}}^{}\right\}^{\frac{1}{2}}_{}\,,\\
[2mm]
\tilde{\varepsilon}_X^{}&=&\frac{1}{2\pi}\frac{\kappa'^2_{}(3\bar{f}'^2_{}
+\bar{\kappa}'^2_{})(2-r^2_{})r^2_{}\sin\delta}{4\left(3\bar{f}'^2_{}+\bar{\kappa}'^2_{}\right)r^2_{}
\left(1+\cos\delta\right)+3\kappa'^2_{}\left(6+r^4_{}\right)
}\,.\nonumber\\
&&
\end{eqnarray}
It is easy to understand that the parameter $y'$ in the reheating
temperature does not appear in the CP asymmetry since the CP
asymmetry is from the interference between the tree and one-loop
diagrams of the three-body decays. So, we can obtain a sizable CP
asymmetry even if the $y'$ is smaller to lower the reheating
temperature. By inputting
\begin{eqnarray}
&&y'=4\times 10^{-6}_{}\,,~~\bar{f}'=\bar{\kappa}'=0.5\,,~~\kappa'=0.04\,,\nonumber\\
&& r=0.1\,,~~\sin\delta= 0.75\,,
\end{eqnarray}
we can induce a reheating temperature as low as
\begin{eqnarray}
T_R^{}\simeq 10^{8}_{}\,\textrm{GeV}
\left(\frac{M_X^{}}{10^{13}_{}\,\textrm{GeV}}\right)^{\frac{1}{2}}_{}\,.
\end{eqnarray}
The constraints from the gravitino number density imply that the
reheating temperature should have an upper bound $\sim
10^{7-10}\,\textrm{GeV}$ \cite{kl1983}. Therefore, our model can
avoid the gravitino problem. Furthermore, we can have a CP asymmetry
\begin{eqnarray}
\tilde{\varepsilon}\simeq 4.0\times 10^{-5}
\end{eqnarray}
to generate the cosmological baryon asymmetry
\begin{eqnarray}
\eta_B^{}\simeq
10^{-10}\left(\frac{M_X^{}}{10^{13}_{}\,\textrm{GeV}}\right)^{-\frac{1}{2}}_{}\,.
\end{eqnarray}

By the exchange of one isosinglet and two isotriplet diquark
scalars, there will be six-quark interactions violating the baryon
number of two units, as shown in Fig. \ref{bviolation}. The
effective operators should be
\begin{eqnarray}
\mathcal{L}_{eff}^{\Delta B
=2}&\supset&-\sum_{a}^{}\frac{4\kappa_a^{}\bar{f}_{aij}^\ast
h_{kl}^\ast h_{mn}^\ast}{M_a^{} M_\Omega^4}\bar{d}_{Ri}^{}d_{Rj}^c
\left(\bar{u}_{Lk}^{}d_{Ll}^c\bar{u}_{Lm}^{}d_{Ln}^c\right.\nonumber\\
&& \left.- \frac{1}{2}
\bar{u}_{Lk}^{}u_{Ll}^c\bar{d}_{Lm}^{}d_{Ln}^c\right)+\textrm{H.c.}\,,
\end{eqnarray}
from which we can read the operators for the neutron-antineutron
oscillation,
\begin{eqnarray}
\label{nbarn}
\mathcal{L}_{eff}^{n-\bar{n}}&=&-\sum_{a}^{}\frac{6\kappa_a^{}\bar{f}_{a11}^\ast
h_{11}^\ast h_{11}^\ast}{M_a^{} M_\Omega^4}\bar{d}_{R}^{}d_{R}^c
\bar{u}_{L}^{}d_{L}^c\bar{u}_{L}^{}d_{L}^c +\textrm{H.c.}\,.\nonumber\\
&&
\end{eqnarray}
With the previous parameter choice for the inflationary
baryogenesis, the induced neutron-antineutron oscillation can arrive
at a testable level \cite{takita1986} as it has a strength of the
order of
\begin{eqnarray}
G^{}_{n-\bar{n}}&\sim&\sum_{a}^{}\frac{6\kappa_a^{}\bar{f}_{a11}^\ast
h_{11}^\ast h_{11}^\ast}{M_a^{} M_\Omega^4}\sim\frac{12 r \kappa'
\bar{f}' h^{2}_{11}}{M_X^{}M_\Omega^4}\nonumber\\
&=& 3\times
10^{-28}_{}\textrm{GeV}^{-5}_{}\left(\frac{r}{0.1}\right)
\left(\frac{\kappa'}{0.5}\right)\left(\frac{\bar{f}'}{0.5}\right)\nonumber\\
&&\times \left(\frac{h_{11}^{}}{0.1}\right)^2_{}
\left(\frac{M_X^{}}{10^{13}_{}\,\textrm{GeV}}\right)^{-1}_{}
\left(\frac{M_\Omega^{}}{1\,\textrm{TeV}}\right)^{-4}_{}\,.
\end{eqnarray}

In this paper we studied a supersymmetric model with the singlet and
diquark superfields. The scalar component of the singlet superfield
can drive a chaotic inflation. The inflaton only have the three-body
decays mediated by the isosinglet diquark scalars and their
superpartners. The inflaton decays can simultaneously allow a low
reheating temperature to avoid the gravitino problem and a sizable
CP asymmetry to generate the cosmological baryon asymmetry. There
will emerge an observable neutron-antineutron oscillation if the
isotriplet diquark scalar is at the TeV scale. The isotriplet
diquark could be verified at colliders such as the LHC
\cite{agms2009}. Furthermore, our model can be modified and
extended. For example, the isotriplet diquark superfields can be
replaced by the isosinglet ones. We can also introduce more singlet
superfields to give other inflationary scenarios, like the double
sneutrino inflation \cite{bfz2003} and the hybrid sneutrino
inflation \cite{abks2005}.

\textbf{Acknowledgement}: PHG is supported by the Alexander von
Humboldt Foundation. He is also indebted to Ngee-Pong Chang and
Kok-Khoo Phua for a warm hospitality at the Nanyang Technological
University in Singapore where part of this work was performed. US
thanks R. Cowsik for arranging his visit as the Clark Way Harrison
visiting professor.

\end{document}